\documentclass[conference,a4paper]{IEEEtran}

\IEEEoverridecommandlockouts 

\usepackage[utf8]{inputenc} 
\usepackage[T1]{fontenc}
\usepackage{url}              
\usepackage{cite}             

\usepackage[cmex10]{amsmath}  
\usepackage{mleftright}       
\mleftright                   

\usepackage{graphicx}         
\usepackage{booktabs}         

\usepackage{ragged2e}
\usepackage{amsfonts}
\usepackage{enumitem}
\usepackage{bm}
\usepackage{amssymb}
\usepackage{makecell}
\usepackage{blindtext}
\usepackage{multicol}
\usepackage{color}
\usepackage{balance}

\newtheorem{definition}{Definition}
\newtheorem{lemma}{Lemma}
\newtheorem{remark}{Remark}
\newtheorem{theorem}{Theorem}
\newtheorem{example}{Example}
\newtheorem{corollary}{Corollary}

\newcommand{\floor}[1]{\left\lfloor #1 \right\rfloor}

\newlist{mycases}{enumerate}{1}
\setlist[mycases]{label=\textit{Case~\Roman*:},align=left,itemindent=0pt,leftmargin=0pt,labelwidth=-4pt}

\begin{document}

\title{Direct Constructions of Multiple Shift Complementary Sets of Flexible Lengths}

 \author{
	   \IEEEauthorblockN{Abhishek Roy}
	   \IEEEauthorblockA{Department of Mathematics\\
		                    Indian Institute of Technology Patna \\
		                   Bihta, Patna, Bihar-801106, India\\
		                    1821ma06@iitp.ac.in}
	   \and
	   \IEEEauthorblockN{Sudhan Majhi}
	  \IEEEauthorblockA{Department of Electrical Communication Engineering\\
		                Indian Institute of Science \\ 
		                 Bangalore, Karnataka-560012, India\\
	                     smajhi@iisc.ac.in}
                     
       \thanks{The work of Abhishek Roy was supported by CSIR-SRF, Government of India, File no. 09/1023(0025)/2018-EMR-I.}          
 }

\maketitle

\begin{abstract}
   Golay complementary set (GCS) plays a vital role in reducing peak-to-mean envelope power ratio (PMEPR) in orthogonal frequency division multiplexing (OFDM). A more general version of GCS is a multiple shift complementary set (MSCS), where by relaxing the condition of zero auto-correlation sum throughout all the non-zero time shifts to the integer multiples of some fixed time shift, more sequence sets can be made available. In this paper, we propose direct constructions of MSCSs with flexible and arbitrary lengths and flexible set sizes, by using multivariable functions, which have not been reported before.
  
\end{abstract}


\section{Introduction}
Golay complementary pair (GCP) was first conceptualized in 1951 by Marcel J. E. Golay \cite{GolayInfra}. It is a pair of sequences which has zero aperiodic auto-correlation function (AACF) sum for all non-zero time shifts.  Because of its ideal AACF property, it has been widely used in orthogonal frequency division multiplexing (OFDM) \cite{OFDM1, OFDM2}, radar \cite{radar}, channel estimation \cite{channel}  etc. In OFDM system, GCP carries out the role of reducing peak-to-mean envelope power ratio (PMEPR) \cite{Davis}. The first direct construction of GCP appears in \cite{Davis}, where $2^h$-ary ($h \geq 1$ is an integer) GCPs were constructed by generalized Boolean functions (GBFs). The idea of GCP was extended to Golay complementary set (GCS) which is a set of more than two sequences having zero AACF sum for all non-zero time shifts \cite{GCS1}. Although there are many constructions of GCS in the literature \cite{schmidt, CSchao, CSchao1, shibuccc1, shibuccc2, palashccc}, Paterson \textit{et al.} first proposed a direct construction of GCS using GBFs \cite{paterson}. Like GCPs, GCSs are also used in OFDM system to reduce PMEPR, where it is upper bounded by the number of sequences in the GCS. 

The multiple shift complementary set (MSCS) is a more general version of GCS, where the AACF sum is zero for multiples of some fixed time shift. It was first introduced by Xin and Fair as an alternative to GCS \cite{xin}. Later Chen \textit{et al.} provided a direct construction of MSCS using GBFs \cite{chenMSCS} with sequence length of the form of power-of-two.  In \cite{MSCS2}, the authors studied even-shift complementary pairs, which is a special case of MSCS, where AACF sum equals zero when the time shift is even. Also, in this case the number of constituent sequences in the set or the set size is $2$. Recently, Chen \textit{et al.} proposed a direct construction of MSCS of non-power-of-two length by using GBFs \cite{ChaoMSCS}. But the length is of the form $2^{m-1}+2^t$, where $m \geq 2$ and $1 \leq t \leq m-1$. To the best of authors' knowledge, there is no direct construction of MSCS of arbitrary lengths and set sizes.

Motivated by this, in this paper, we propose constructions of MSCSs with flexible and arbitrary lengths by using multivariable functions. The lengths of the proposed MSCSs are of the form $p_1^{m_1} p_2^{m_2} \dots p_k^{m_k} $ and $p_1^{m_1} p_2^{m_2} \dots p_k^{m_k} p_{k+1}$, where $p_i$'s are prime numbers and $m_i \geq 1$ are integers $\forall i=1,2,\dots,k$. The set sizes of the MSCSs are of the form $p_1 p_2 \dots p_k$ and $p_{k+1}$. 

The rest of the paper is organized as follows. In Section II, some definitions are provided. Later in Section III, the main construction of the MSCSs are proposed. The PMEPRs of the proposed constructions are also investigated in this section. Finally, in Section IV, concluding remarks have been given.

\section{Preliminaries}

\begin{definition}
	Let $\mathbf{a}= (a_0, a_1, \dots, a_{L-1})$ and $\mathbf{b}= (b_0, b_1, \dots, b_{L-1})$ be two complex-valued sequence of length $L$. Then the aperiodic cross-correlation function (ACCF) at time shift $\tau$ is defined by
	\begin{equation}
		\rho(\mathbf{a}, \mathbf{b}) (\tau)=
		\begin{cases}
			\sum_{i=0}^{L-1-\tau} a_i b^*_{i+\tau} ,& 0 \leq \tau < L;\\
			\sum_{i=0}^{L-1+\tau} a_{i-\tau} b^*_{i} ,& -L < \tau < 0,\\
		\end{cases}
	\end{equation}
where $(\cdot)^*$ denotes the complex conjugate. When $\mathbf{a}=\mathbf{b}$, then it is called AACF and denoted by $\rho(\mathbf{a})(\tau)$.
\end{definition}

\begin{definition}[GCS]
A set of sequences $\{\mathbf{a}_0, \mathbf{a}_1, \dots, \mathbf{a}_{M-1}\}$ with length $L$ is called a GCS if they satisfy 
\begin{equation}
	\sum_{i=0}^{M-1} \rho(\mathbf{a}_i) (\tau) =0, \forall \tau \neq 0.
\end{equation}

If $M=2$, then it is called a GCP. 
\end{definition}

\begin{definition}[MSCS]
	A set of sequences $\{\mathbf{a}_0, \mathbf{a}_1, \dots, \mathbf{a}_{M-1}\}$ with length $L$ is called a MSCS if for some positive number $S$ they satisfy 
	\begin{equation}
	\sum_{i=0}^{M-1} \rho(\mathbf{a}_i) (\tau) =0,  \tau \neq 0 ~\&~ \tau ~\text{mod}~ S=0.
 \end{equation}
It is denoted by $(M,L,S)$-MSCS, where $M$ is called the set size. It should be noted that when $S=1$, then it becomes a GCS.
\end{definition}

\begin{definition}[Type-II ZCS]
  A set of sequences $\{\mathbf{a}_0, \mathbf{a}_1, \dots, \mathbf{a}_{M-1}\}$ with length $L$ is called a type-II Z-complementary set (ZCS) if for some positive number $Z$ they satisfy 
  \begin{equation}
  	\sum_{i=0}^{M-1} \rho(\mathbf{a}_i) (\tau) =0,  L-Z< \left|  \tau \right| <L.
  \end{equation}
Here $Z$ is called the zero correlation zone (ZCZ) width. The set may be denoted as type-II $(M,L,Z)$-ZCS.
\end{definition}

\subsection{Multivariable Function}
Let $\mathbb{Z}_{p}= \{0,1, \dots, p-1\}$ be the set of integers modulo $p$. A multivariable function can be defined as $$f: \mathbb{Z}_{p_1}^{m_1} \times \mathbb{Z}_{p_2}^{m_2} \times \dots \times \mathbb{Z}_{p_k}^{m_k} \rightarrow \mathbb{Z}_{\lambda}  $$
where $p_1, p_2, \dots, p_k$ are prime numbers, $m_i \geq 1$, $\forall i=1,2, \dots, k$ and $\lambda$ is a positive integer. $v_{p_\alpha,1}, v_{p_\alpha,2}, \dots, v_{p_\alpha, m_\alpha}$ are the $m_\alpha$ variables which takes values from $\mathbb{Z}_{p_\alpha}$ for $\alpha=1,2, \dots, k$. The set of monomials of degree at most $r$ is given by $A^r= \left\{\prod_{\alpha=1}^{k} \prod_{\beta=1}^{m_\alpha} v_{p_\alpha, \beta}^{j_\beta^\alpha} : 0 \leq \sum_{\alpha=1}^{k} \sum_{\beta=1}^{m_\alpha} j_\beta^\alpha \leq r \right\}$. A multivariable function of order $r$ is a $\mathbb{Z}_\lambda$-linear combination of the monomials from $A^r$. Let $\mathbf{u}_{p_\alpha, i_\alpha}$ be the vector representation of $i_\alpha$ with base $p_\alpha$, i.e., $ \mathbf{u}_{p_\alpha, i_\alpha}= (i_{\alpha,1}, i_{\alpha,2}, \dots, i_{\alpha,m_\alpha})$ where $i_\alpha= \sum_{\gamma=1}^{m_\alpha} p_\alpha^{\gamma-1} i_{\alpha, \gamma}$. The function rule for a multivariable function is defined as 
$(\mathbf{u}_{p_1,i_1}, \mathbf{u}_{p_2,i_2}, \dots, \mathbf{u}_{p_k,i_k}  ) \longmapsto f(\mathbf{u}_{p_1,i_1}, \mathbf{u}_{p_2,i_2}, \dots, \mathbf{u}_{p_k,i_k}  ) \mod \lambda$. One can associate a $\mathbb{Z}_\lambda$-valued sequence of length $p_1^{m_1} p_2^{m_2} \dots p_k^{m_k}$ corresponding to a multivariable function $f$ as
$	\mathbf{f}= \bigg( f( \mathbf{u}_{p_1,i_1}, \mathbf{u}_{p_2,i_2}, \dots, \mathbf{u}_{p_k,i_k}   ):  i_1=0, 1, \dots, p_1^{m_1}-1;  \dots ; i_k=0, 1, \dots, p_k^{m_k}-1 \bigg)$. Also, one can associate a complex-valued sequence of length $p_1^{m_1} p_2^{m_2} \dots p_k^{m_k}$ in a similar manner as $\psi(\mathbf{f})= \bigg(\omega_\lambda^{ f( \mathbf{u}_{p_1,i_1}, \mathbf{u}_{p_2,i_2}, \dots, \mathbf{u}_{p_k,i_k}   )}:  i_1=0, 1, \dots, p_1^{m_1}-1; \dots ; i_k=0, 1, \dots, p_k^{m_k}-1 \bigg)$, where $\omega_\lambda= \exp(2 \pi \sqrt{-1}/\lambda)$.

.


\section{Construction}

In this section, we propose constructions of MSCSs by using multivariable functions.  
\begin{theorem}\label{thm_main}
	Let $\pi$ be a permutation on the set $\{s, s+1, \dots, m\}$ for integers  $m \geq 1$ and  $s \geq 1$. Let $p$ be a prime and $\lambda$ be a positive integer such that $p \mid \lambda$. Let $f : \mathbb{Z}_{p}^m \rightarrow \mathbb{Z}_{\lambda}$ be defined such that 
	\begin{equation}
		\begin{split}
			f(v_1, v_2, \dots, v_m)= &\frac{\lambda}{p} \sum_{i=s}^{m-1} v_{\pi(i)} v_{\pi(i+1)}  + \sum_{i=1}^{m} g_i v_i  + g\\
			&+ h(v_1, v_2, \dots, v_{s-1}),
		\end{split}
	\end{equation}
	where $g_i, g \in \mathbb{Z}_\lambda$ and $h(v_1, v_2, \dots, v_{s-1})$ is any function $h: \mathbb{Z}_{p}^{s-1} \rightarrow \mathbb{Z}_\lambda$. For $s=1$, we define $h=0$. Define $a^\gamma: \mathbb{Z}_p^m \rightarrow \mathbb{Z}_\lambda$ such that 
	$$a^\gamma= f(v_1, v_2, \dots, v_m)+ \frac{\lambda}{p} v_{\pi(s)} \gamma.$$ 
	Then $\{\omega_\lambda^{a^\gamma} : \gamma \in \mathbb{Z}_p\}$ is a $(p,p^m, p^{s-1})$-MSCS.
\end{theorem}

\begin{IEEEproof}
	As, $\rho(\mathbf{a}) (-\tau)= \rho^*(\mathbf{a}) (\tau)$, we shall only prove for  $\tau \geq 0$. We have to show that
	\begin{equation}
		\sum_{\gamma=0}^{p-1} \sum_{i=0}^{p^m-1-\tau} \omega_\lambda^{ (a^\gamma)_i- (a^\gamma)_{i+\tau}} =0,
	\end{equation}
whenever $\tau \mod p^{s-1}=0$, where $(a^\gamma)_i= a^\gamma(i_1, i_2, \dots, i_m)$, and $(i_1, i_2, \dots, i_m)$ is the $p$-ary vector representation of $i$. Let $j=i+\tau$, where $\tau \mod p^{s-1}=0$, i.e., $\tau$ is a multiple of $p^{s-1}$. 
	We have
	\begin{equation}
		 (a^\gamma)_i- (a^\gamma)_j= \left( f_i-f_j\right)+ \frac{\lambda}{p} \left( i_{\pi(s)} - j_{\pi(s)}\right) \gamma.
	\end{equation}
	Now, we have two cases.
	\begin{mycases}
		\item  Let, $i_{\pi(s)} \neq j_{\pi(s)}$. In this case, we have 
		\begin{equation}
		\sum_{\gamma=0}^{p-1}	\omega_\lambda^{\frac{\lambda}{p} (i_{\pi(s)} - j_{\pi(s)}) \gamma}= \sum_{\gamma=0}^{p-1}  \omega_p^{(i_{\pi(s)} - j_{\pi(s)}) \gamma }=0,
		\end{equation}
	as $ \omega_p^{(i_{\pi(s)} - j_{\pi(s)}) \gamma }$ are the $p$-th roots of $1$ for $\gamma=0,1, \dots, p-1$. So, we have
	\begin{equation}
		\sum_{\gamma=0}^{p-1} \omega_\lambda^{(a^\gamma)_i- (a^\gamma)_j}=0.
	\end{equation}
	
	\item  Let, $i_{\pi(s)} = j_{\pi(s)}$. But $i \neq j$ and $j=i +\tau$, where $\tau$ is a multiple of $p^{s-1}$, i.e., $\tau= k p^{s-1}$ for some $1 \leq k \leq p^{m-s+1}-1$. Any $k$ in this range can be written as a $(m-s+1)$-tuple vector representation form $(k_1, k_2, \dots, k_{m-s+1})$ with base $p$, where $k= \sum_{i=1}^{m-s+1} k_i p^{i-1}$. So, $\tau= \sum_{i=1}^{m-s+1} k_i p^{i+s-2}$, which implies $i_l = j_l$ for $l=1,2, \dots, s-1$. Now, $i \neq j$ implies that $\exists$ some $l \in \{s, s+1, \dots, m\}$ such that $i_l \neq j_l$. Let $\phi$ be the smallest number such that $i_{\pi(\phi)} \neq j_{\pi(\phi)}$. Let $i^\delta$ be the integer whose $p$-ary vector representation is 
	$$(i_1, i_2, \dots, i_{\pi(\phi-1)} -\delta, \dots, i_m), ~\text{if}~ i_{\pi(\phi-1)} -\delta \geq 0$$
and
	 $$(i_1, i_2, \dots, p+  i_{\pi(\phi-1)}-\delta, \dots, i_m),  ~\text{if}~ i_{\pi(\phi-1)} -\delta < 0,$$
	where $\delta \in \{1,2, \dots, p-1\}$ and it differs form the $p$-ary vector representation of $i$ only in the $\pi(\phi-1)$-th position. Similarly, we take $j^\delta$.
		Now, for $i_{\pi(\phi-1)}-\delta \geq 0$ and $j_{\pi(\phi-1)}-\delta \geq 0$, we have 
		\begin{equation}
			(a^\gamma)_{i^\delta}- 	(a^\gamma)_{i} = -\delta \left(\frac{\lambda}{p} i_{\pi(\phi-2)} + \frac{\lambda}{p} i_{\pi(\phi)}  + g_{\pi(\phi-1)}\right) 
		\end{equation}
	and
	\begin{equation}
		(a^\gamma)_{j^\delta}- 	(a^\gamma)_{j} =  -\delta \left(\frac{\lambda}{p} j_{\pi(\phi-2)} + \frac{\lambda}{p} j_{\pi(\phi)}  + g_{\pi(\phi-1)}\right).
	\end{equation}
But, we have
\begin{equation}
	\begin{split}
		&\left((a^\gamma)_{i^\delta}- 	(a^\gamma)_{j^\delta}\right)- 	\left((a^\gamma)_{i}- 	(a^\gamma)_{j}\right)\\
		&= \left(	(a^\gamma)_{i^\delta}- 	(a^\gamma)_{i}\right)- \left((a^\gamma)_{j^\delta}- 	(a^\gamma)_{j}\right) \\
		&= -\delta \frac{\lambda}{p} \left( i_{\pi(\phi)} - j_{\pi(\phi)}\right).
	\end{split}
\end{equation}
Also, for $i_{\pi(\phi-1)}-\delta < 0$ and $j_{\pi(\phi-1)}-\delta <0$, we have 
\begin{equation}
	\begin{split}
		&\left((a^\gamma)_{i^\delta}- 	(a^\gamma)_{j^\delta}\right)- 	\left((a^\gamma)_{i}- 	(a^\gamma)_{j}\right)\\
		&= (p-\delta) \frac{\lambda}{p} \left( i_{\pi(\phi)} - j_{\pi(\phi)}\right).
	\end{split}
\end{equation}
But $\omega_\lambda^{(p-\delta) \frac{\lambda}{p} \left( i_{\pi(\phi)} - j_{\pi(\phi)}\right)}= \omega_\lambda^{-\delta \frac{\lambda}{p} \left( i_{\pi(\phi)} - j_{\pi(\phi)}\right)}$.
So, considering all the possibilities, we have
\begin{equation}
	\begin{split}
		&\sum_{\delta=1}^{p-1} \omega_\lambda^{\left((a^\gamma)_{i^\delta}- 	(a^\gamma)_{j^\delta} \right)- 	\left((a^\gamma)_{i}- 	(a^\gamma)_{j}\right)}= \sum_{\delta=1}^{p-1} \omega_p^{\delta (j_{\pi(\phi)} - i_{\pi(\phi)})}\\
		& \implies \sum_{\delta=1}^{p-1} \omega_\lambda^{\left((a^\gamma)_{i^\delta}- 	(a^\gamma)_{j^\delta} \right)- 	\left((a^\gamma)_{i}- 	(a^\gamma)_{j}\right)}= -1\\
		& \implies \sum_{\delta=1}^{p-1} \omega_\lambda^{\left((a^\gamma)_{i^\delta}- 	(a^\gamma)_{j^\delta} \right)} + \omega_\lambda^{\left((a^\gamma)_{i}- 	(a^\gamma)_{j}\right)}=0.
	\end{split}
\end{equation}
Hence, we have the result.
	\end{mycases}
\end{IEEEproof}

In the following example, we illustrate the \textit{Theorem \ref{thm_main}}.
\begin{example}\label{eg_thm_main}
	Let, $p=3$, $m=3$, $s=2$, $q=6$ and $\pi$ be a permutation on $\{2,3\}$ such that $\pi(2)=2$, $\pi(3)=3$. From \textit{Theorem \ref{thm_main}}, we can construct the function $f : \mathbb{Z}_3^3 \rightarrow \mathbb{Z}_6$, where $f(v_1, v_2, v_3)= 2 v_2 v_3 + 5$. Then $\{\omega_3^{a^\gamma} : \gamma \in \mathbb{Z}_3\}$ is a $(3,27,3)$-MSCS, where $a^\gamma= f+ 2v_2 \gamma$.
\end{example}

We can show that the set $\{\omega_\lambda^{a^\gamma}: \gamma=0,1,\dots, p-1\}$ constructed in \textit{Theorem \ref{thm_main}} has some interesting property which is pretty straightforward and we show it in the next corollary.

\begin{corollary}\label{cor_1}
	Let $a^\gamma$ be the function defined in \textit{Theorem \ref{thm_main}}. Then we have
	\begin{equation}
		\sum_{\gamma=0}^{p-1} \rho(\omega_\lambda^{a^\gamma}) (\tau)= 0,
	\end{equation}
	when $\left| \tau \right|> p^{s-1} $.
\end{corollary}

\begin{IEEEproof}
	We shall prove for $\tau >0$ as $\rho(\mathbf{a}) (-\tau)= \rho^*(\mathbf{a}) (\tau)$. For $\tau > p^{s-1}$, let $j=i+\tau$. As $\tau > p^{s-1}$, it is not possible that  $i_k = j_k, \forall k \in \{s, s+1, \dots, m\}$.
	Now, $(a^\gamma)_i-(a^\gamma)_j= (f_i -f_j)+\frac{\lambda}{p} \left( i_{\pi(s)} - j_{\pi(s)}\right) \gamma$. So, if $i_{\pi(s)} \neq j_{\pi(s)}$, then we have 
	\begin{equation}
		\sum_{\gamma=0}^{p-1} \omega_\lambda^{(a^\gamma)_i-(a^\gamma)_j}= \omega_\lambda^{f_i-f_j} \sum_{\gamma=0}^{p-1} \omega_\lambda^{\frac{\lambda}{p} \left( i_{\pi(s)} - j_{\pi(s)}\right) \gamma}=0.
	\end{equation}
	If $i_{\pi(s)} = j_{\pi(s)}$, then in a similar manner described in \textit{Theorem \ref{thm_main}}, we find $i^\delta$ and $j^\delta$ for $\delta=1,2, \dots, p-1$. 
	Now, arguing similar to \textit{Theorem \ref{thm_main}}, we have 
	\begin{equation}
		\sum_{\gamma=0}^{p-1} \left( \sum_{\delta=1}^{p-1} \omega_\lambda^{\left((a^\gamma)_{i^\delta}- 	(a^\gamma)_{j^\delta} \right)} + \omega_\lambda^{\left((a^\gamma)_{i}- 	(a^\gamma)_{j}\right)} \right)=0.
	\end{equation}
	Hence, the result.
\end{IEEEproof}

\begin{remark}
	\textit{Corollary \ref{cor_1}} shows that the set $\{a^{\gamma}: \gamma=0,1,\dots, p-1\}$ is in fact a type-II $(p, p^m, p^m-p^{s-1})$-ZCS. Type-II ZCS has application in wireless communication. For example, it can be applied in  wideband wireless communication system having large minimum interfering signal delay (ISD) for removing asynchronous interference \cite{type2appli, rajenType2ZCP}. Type-II ZCSs are preferable than type-I ZCSs in these scenarios to mitigate inter-symbol interference \cite{mobilecom}. Although, there are some  indirect \cite{rajentype2zccs} and direct \cite{rajenType2ZCP} constructions of type-II ZCS in the literature, the proposed construction is direct, as well as provides flexible ZCZ width when the length is power-of-prime. 
\end{remark}

Next, we generalize the \textit{Theorem \ref{thm_main}} for MSCS. 
\begin{theorem}\label{thm_gen}
	Let $p_1, p_2, \dots, p_k$ be $k$ primes and $\lambda$ be a positive integer such that $p_\alpha \mid \lambda, \forall \alpha=1,2, \dots, k;$ and $\pi_{\alpha}$ be permutations on the sets $I_\alpha= \{s_{\alpha}, s_{\alpha}+1, \dots, m_{\alpha}\}$ for $s_\alpha \geq 1$ and $m_\alpha \geq 1$, where $\alpha=1,2, \dots, k$. Let $f_\alpha: \mathbb{Z}_{p_\alpha}^{m_\alpha} \rightarrow \mathbb{Z}_{\lambda}$ be defined by
	\begin{equation}
		\begin{split}
				&f_\alpha(v_{p_\alpha,1}, v_{p_\alpha,2}, \dots, v_{p_\alpha,m_\alpha})\\
				&= \frac{\lambda}{p_\alpha} \sum_{i=s_\alpha}^{m_\alpha-1} v_{p_\alpha,\pi_\alpha(i)} v_{p_\alpha,\pi_\alpha(i+1)}  + \sum_{i=s_\alpha}^{m_\alpha} g_{p_\alpha, i} v_{p_\alpha,i} + g_{p_\alpha}\\
				&~~~~+ h_\alpha (v_{p_\alpha,1} v_{p_\alpha,2}, \dots, v_{p_\alpha,s_\alpha-1}),
		\end{split}
	\end{equation} 
	where $g_{p_\alpha,i}, g_{p_\alpha} \in \mathbb{Z}_\lambda$, $h_\alpha (v_{p_\alpha,1} v_{p_\alpha,2}, \dots, v_{p_\alpha,s_\alpha-1})$ is any function $h_\alpha: \mathbb{Z}_{p_\alpha}^{s_\alpha-1} \rightarrow \mathbb{Z}_\lambda$ and $h_\alpha=0$ when $s_\alpha=1$. We define $$a^{\bm{\gamma}} : \mathbb{Z}_{p_1}^{m_1} \times \mathbb{Z}_{p_2}^{m_2} \times \dots \times \mathbb{Z}_{p_k}^{m_k} \rightarrow \mathbb{Z}_{\lambda},$$
	for $\bm{\gamma}=(\gamma_1, \gamma_2, \dots, \gamma_k) \in \mathbb{Z}_{p_1} \times \mathbb{Z}_{p_2} \times \dots \times \mathbb{Z}_{p_k}$ such that
	\begin{equation}
		\begin{split}
			a^{\bm{\gamma}} = \sum_{\alpha=1}^{k} f_\alpha + \sum_{\alpha=1}^{k} \frac{\lambda}{p_\alpha} v_{p_\alpha, \pi_\alpha(s_\alpha)} \gamma_\alpha.
		\end{split}
	\end{equation}
Then the set $\{\omega_\lambda^{a^{\bm{\gamma}} }: \bm{\gamma} \in  \mathbb{Z}_{p_1} \times \mathbb{Z}_{p_2} \times \dots \times \mathbb{Z}_{p_k}\}$ is a $\left(\prod_{\alpha=1}^{k} p_\alpha, \prod_{\alpha=1}^{k} p_\alpha^{m_\alpha}, \prod_{\alpha=1}^{k} p_\alpha^{s_\alpha-1} \right)$-MSCS.
\end{theorem}

\begin{table*}[t]
	\centering
	\caption{Comparison Table for Parameters}
	\begin{tabular}{ |c| c| c| c| c| c|}
		\hline
		Construction & Method &Length & \makecell{Set  Size} & \makecell{Value of $S$}& Constraints\\ 
		\hline
		\hline
		\cite{chenMSCS} & GBF &$2^m$ & $2$ & $2^d$ & \makecell{$m \geq1$, $0 \leq d < m$} \\  
		\hline
		\cite{MSCS2} &  Matrix operation &  \makecell{$2N+1$, $2N+2$, \\ $2 N_1$, $N_1 N$,\\ $N_1 + N_2$ }& $2$ & $2$ & \makecell{$N$ is length of existing GCP, $N_1$ and $N_2$ are lengths\\  of existing MSCS} \\  
		\hline
		\cite{ChaoMSCS} & GBF & $2^{m-1}+ 2^t$ & $2^{k-s+1}$& $2^s$& \makecell{$m \geq 2$, $1 \leq t \leq m-1$,  $k \leq m-1$, $1 \leq s \leq k' \leq k$} \\  
		\hline
		\makecell{Proposed,\\ \textit{Theorem \ref{thm_gen}} }&  \makecell{Multivariable\\ function }& $\prod_{\alpha=1}^{k} p_\alpha^{m_\alpha} $ & $\prod_{\alpha=1}^{k} p_\alpha$ &$\prod_{\alpha=1}^{k} p_\alpha^{s_\alpha-1}$  & \makecell{$p_\alpha$'s are prime $\forall \alpha$, $m_\alpha \geq 1, \forall \alpha$,\\ $1 \leq s_\alpha \leq m_\alpha$, $\forall \alpha$. } \\  
		\hline
		\makecell{Proposed,\\ \textit{Theorem \ref{thm_gen_2}} }&  \makecell{Multivariable\\ function }& $ p_{k+1} \prod_{\alpha=1}^{k} p_\alpha^{m_\alpha}$& $\prod_{\alpha=1}^{k} p_\alpha$ & $ p_{k+1}$ & \makecell{$p_\alpha$'s are prime $\forall \alpha$, $m_\alpha \geq 1, \forall \alpha=1,2,\dots,k$,\\ $ s_\alpha =1$, $\forall \alpha=1,2,\dots,k$. } \\  
		\hline
	\end{tabular}
	\label{table_com}
\end{table*}

\begin{IEEEproof}
	We prove this by induction. From \textit{Theorem \ref{thm_main}}, it is evident that the statement is true for $k=1$. We assume that the statement is true for $k=n$. Then we have to prove that it is true for $k=n+1$. For $k=n+1$, we have 
	\begin{equation}
		\begin{split}
				a^{\bm{\gamma}} = & \sum_{\alpha=1}^{n+1} f_\alpha + \sum_{\alpha=1}^{n+1} \frac{\lambda}{p_\alpha} v_{p_\alpha, \pi_\alpha(s_\alpha)} \gamma_\alpha\\
				=& \left( \sum_{\alpha=1}^{n} f_\alpha + \sum_{\alpha=1}^{n} \frac{\lambda}{p_\alpha} v_{p_\alpha, \pi_\alpha(s_\alpha)} \gamma_\alpha \right)\\
				& + \left( f_{n+1} + \frac{\lambda}{p_{n+1}} v_{p_{n+1}, \pi_{n+1} (s_{n+1}) \gamma_{n+1}}\right)\\
				=& R +S
		\end{split}
	\end{equation}
where
\begin{equation}
	R=  \sum_{\alpha=1}^{n} f_\alpha + \sum_{\alpha=1}^{n} \frac{\lambda}{p_\alpha} v_{p_\alpha, \pi_\alpha(s_\alpha)} \gamma_\alpha,
\end{equation}
and
\begin{equation}
	S= f_{n+1} + \frac{\lambda}{p_{n+1}} v_{p_{n+1}, \pi_{n+1} (s_{n+1}) \gamma_{n+1}}.
\end{equation}
Now, $\omega_\lambda^{a^{\bm{\gamma}}}= \omega_\lambda^{S} \otimes \omega_\lambda^{R}$, where $\otimes$ denotes the Kronecker product. It is easy to observe that the length of each sequence in $\{\omega_\lambda^{a^{\bm{\gamma}} }: \bm{\gamma} \in  \mathbb{Z}_{p_1} \times \mathbb{Z}_{p_2} \times \dots \times \mathbb{Z}_{p_k}\}$ is $L= \prod_{\alpha=1}^{n+1} p_{\alpha}^{m_\alpha}$ and number of sequence is $M= \prod_{\alpha=1}^{n+1} p_{\alpha}$. We need to show that 
\begin{equation}
\sum_{\bm{\gamma} \in \mathbb{Z}_{p_1} \times \mathbb{Z}_{p_2} \times \dots \times \mathbb{Z}_{p_{n+1}}}^{} \rho(\omega_\lambda^{a^{\bm{\gamma}}}) (\tau)=0
\end{equation}
when $\tau \mod \prod_{\alpha=1}^{n+1} p^{s_\alpha-1} =0$. We note that for two sequences $\mathbf{a}$ and $\mathbf{b}$ having length $L_1$ and $L_2$, respectively, we have 
\begin{equation}\label{eqn_kronecker}
	\begin{split}
		 \rho(\mathbf{a} \otimes \mathbf{b}) (\tau)= &\rho(\mathbf{a}) \left( \floor{\frac{\tau}{L_2}}\right) \rho(\mathbf{b}) \left( \tau \mod L_2\right)\\
		 +& \Delta_{L_2} \rho(\mathbf{a}) \left( \floor{\frac{\tau}{L_2}} +1\right) \rho(\mathbf{b}) \left( \tau \!\!\!\!\!  \mod L_2 -L_2\right),\\
	\end{split}
\end{equation}
where
\begin{equation}
	\Delta_{L_2}=
	\begin{cases}
		0, & \tau \mod L_2=0;\\
		1, & \text{otherwise},
	\end{cases}
\end{equation}
and $\floor{\cdot}$ is the floor function.
So, we can write
\begin{equation}\label{eqn_cor_main}
	\begin{split}
		&\sum_{\bm{\gamma} \in \mathbb{Z}_{p_1} \times \mathbb{Z}_{p_2} \times \dots \times \mathbb{Z}_{p_{n+1}}}^{} \rho(\omega_\lambda^{a^{\bm{\gamma}}}) (\tau)\\
		&=\bigg[ \sum_{\gamma_{n+1} \in \mathbb{Z}_{p_{n+1}}}^{} \rho(\omega_\lambda^{S}) \left( \floor{\frac{\tau}{L_2}}\right) \\
		& ~~~~\times \sum_{\bm{\gamma}' \in \mathbb{Z}_{p_1} \times \dots\times \mathbb{Z}_{p_n}}^{} \rho(\omega_\lambda^{R}) \left( \tau \mod L_2\right) \bigg]\\
		&~~~~+ \bigg[ \Delta_{L_2} \times   \sum_{\gamma_{n+1} \in \mathbb{Z}_{p_{n+1}}}^{}    \rho(\omega_\lambda^{S}) \left( \floor{\frac{\tau}{L_2}} +1\right)\\
		& ~~~~\times \sum_{\bm{\gamma}' \in \mathbb{Z}_{p_1} \times \dots\times \mathbb{Z}_{p_n}}^{} \rho(\omega_\lambda^{R}) \left( \tau \!\!\!\!\!  \mod L_2 -L_2\right) \bigg]\\
	\end{split}
\end{equation}
where $\bm{\gamma}= (\bm{\gamma}', \gamma_{n+1})$,$\bm{\gamma}'= (\gamma_1, \gamma_2, \dots, \gamma_n)$ and $L_2= \prod_{\alpha=1}^{n} p_\alpha^{m_\alpha}$.
Now, let $\tau \mod \prod_{\alpha=1}^{n+1} p_\alpha^{s_\alpha-1}=0$, i.e., $\tau = \beta \prod_{\alpha=1}^{n+1} p_\alpha^{s_\alpha-1}$  for some $\beta \in \{1, 2, \dots, \prod_{\alpha=1}^{n+1} p_{\alpha}^{m_\alpha-s_\alpha+1}-1\}$. But either $\beta= \beta_0 \prod_{\alpha=1}^{n} p_{\alpha}^{m_\alpha-s_\alpha+1} $ for some integer $\beta_0$, or $\prod_{\alpha=1}^{n} p_{\alpha}^{m_\alpha-s_\alpha+1} \nmid \beta$. 
Now, we have two cases.
\begin{mycases}
	\item  Let, $\beta= \beta_0 \prod_{\alpha=1}^{n} p_{\alpha}^{m_\alpha-s_\alpha+1} $ for some integer $\beta_0$. Then $\tau \mod L_2=0$ and $\Delta_{L_2}=0$. Also, in this case, $\floor{\frac{\tau}{L_2}}= \beta_0 p_{n+1}^{s_{n+1}-1}$, i.e.,  $\floor{\frac{\tau}{L_2}} \mod p_{n+1}^{s_{n+1}-1}=0$. So, from our assumption, we have 
	\begin{equation}
		\sum_{\gamma_{n+1} \in \mathbb{Z}_{p_{n+1}}}^{} \rho(\omega_\lambda^{S}) \left( \floor{\frac{\tau}{L_2}}\right) =0.
	\end{equation}

\item Let, $\prod_{\alpha=1}^{n} p_{\alpha}^{m_\alpha-s_\alpha+1}$ does not divide $\beta$. Then $\tau \mod L_2 \neq 0$ and $\Delta_{L_2}=1$. Let $\tau \mod L_2=\tau_1$, where $\tau_1= \tau - \beta_1 L_2$ for some integer $\beta_1$ and $0< \tau_1 < L_2$. But $ \prod_{\alpha=1}^{n} p_\alpha^{s_\alpha-1}$ divides both $\tau$ and $L_2$. Hence $\tau_1 \mod \prod_{\alpha=1}^{n} p_\alpha^{s_\alpha-1} =0$, i.e., $\left(\tau \mod L_2\right) \mod \prod_{\alpha=1}^{n} p_\alpha^{s_\alpha-1}=0$. But from our assumption, we have
\begin{equation}\label{eqn_cor_1}
	 \sum_{\bm{\gamma}' \in \mathbb{Z}_{p_1} \times \dots\times \mathbb{Z}_{p_n}}^{} \rho(\omega_\lambda^{R}) \left( \tau \mod L_2\right) =0.
\end{equation}
Arguing in a similar manner, we can say 
\begin{equation}\label{eqn_cor_2}
	\sum_{\bm{\gamma}' \in \mathbb{Z}_{p_1} \times \dots\times \mathbb{Z}_{p_n}}^{} \rho(\omega_\lambda^{R}) \left( \tau \!\!\!\!\!  \mod L_2 -L_2\right)=0.
\end{equation}
\end{mycases}
Hence, the case for $k=n+1$ is proved and we have the result.
\end{IEEEproof}

\begin{remark}\label{remark_1}
	If $s_\alpha=1, \forall \alpha=1,2, \dots, k$, then the set $\{\omega_\lambda^{a^{\bm{\gamma}} }: \bm{\gamma} \in  \mathbb{Z}_{p_1} \times \mathbb{Z}_{p_2} \times \dots \times \mathbb{Z}_{p_k}\}$ becomes a GCS of length $L=p_1^{m_1} p_2^{m_2} \dots p_k^{m_k}$ and set size $M= p_1 p_2 \dots p_k$, which is given in \cite{palashccc}. So, the construction of GCS  in \cite{palashccc} is a special case of the proposed construction.
\end{remark}

	Now, we propose another construction of MSCS.
\begin{theorem}\label{thm_gen_2}
	Let $p_1, p_2, \dots, p_k, p_{k+1}$ be $(k+1)$ distinct primes and $\lambda$ be a positive integer such that $p_\alpha | \lambda, \forall \alpha$. Let $a^{\bm{\gamma}} : \mathbb{Z}_{p_1}^{m_1} \times \mathbb{Z}_{p_2}^{m_2} \times \dots \times \mathbb{Z}_{p_k}^{m_k} \rightarrow \mathbb{Z}_{\lambda}$ be the same function given in \textit{Theorem \ref{thm_gen}} for $s_\alpha=1, \forall \alpha= \{1,2,\dots, k\}$. Then the set $\{b^{\bm{\gamma}}:  \bm{\gamma} \in  \mathbb{Z}_{p_1} \times \mathbb{Z}_{p_2} \times \dots \times \mathbb{Z}_{p_k} \}$ is an $(\prod_{\alpha=1}^{k} p_\alpha, p_{k+1} \prod_{\alpha=1}^{k} p_\alpha^{m_\alpha}, p_{k+1})$-MSCS, where $b^{\bm{\gamma}} : \mathbb{Z}_{p_1}^{m_1} \times \mathbb{Z}_{p_2}^{m_2} \times \dots \times \mathbb{Z}_{p_{k+1}}^{m_{k+1}} \rightarrow \mathbb{Z}_{\lambda}$ is defined by $b^{\bm{\gamma}}= a^{\bm{\gamma}}+ g_{p_{k+1},1} v_{p_{k+1},1} + g_{p_{k+1}} $, where $g_{p_{k+1},1}, g_{p_{k+1}}  \in \mathbb{Z}_\lambda $.
\end{theorem}

\begin{IEEEproof}
	The proof follows form the fact that $b^{\bm{\gamma}}= f_{p_{k+1}} \otimes a^{\bm{\gamma}} $, where $ f_{p_{k+1}}= g_{p_{k+1},1} v_{p_{k+1},1} + g_{p_{k+1}} $, and using (\ref{eqn_kronecker}) and \textit{Remark \ref{remark_1}}, we can get 
	\begin{equation}
		\begin{split}
			\sum_{\bm{\gamma} \in \mathbb{Z}_{p_1} \times \mathbb{Z}_{p_2} \times \dots \times \mathbb{Z}_{p_{k}}}^{} \rho(\omega_\lambda^{a^{\bm{\gamma}}}) (\tau \mod L_2)=&0,\\
				\sum_{\bm{\gamma} \in \mathbb{Z}_{p_1} \times \mathbb{Z}_{p_2} \times \dots \times \mathbb{Z}_{p_{k}}}^{} \rho(\omega_\lambda^{a^{\bm{\gamma}}}) (\tau \mod L_2- L_2)=&0,\\
		\end{split}
	\end{equation}
when $\tau \mod p_{k+1}=0$ and $L_2= \prod_{\alpha=1}^{k} p_\alpha^{m_\alpha}$.
\end{IEEEproof}

\subsection{Peak-to-Mean Envelope Power Ratio (PMEPR)}

Let $\mathbf{x}= (x_0, x_1, \dots, x_{L-1})$ be a $\mathbb{Z}_\lambda$-valued sequence. Then the OFDM signal is the real part of the complex envelope
\begin{equation}
	P_{\mathbf{x}}(t)= \sum_{i=0}^{L-1} \omega_\lambda^{x_i+ \lambda f_i t}
\end{equation}
where $f_i=f+ i \Delta f $, $f$ is a constant frequency, $\Delta f$ is a integer multiple of OFDM symbol rate. The term $\frac{\left| P_{\mathbf{x}} (t) \right|^2}{L}$ is called instantaneous-to-average power ratio (IAPR). PMEPR of the sequence $\mathbf{x}$ is defined as
\begin{equation}
	\text{PMEPR}(\mathbf{x})= \sup_{0 \leq \Delta f t \leq 1} \frac{\left| P_{\mathbf{x}} (t) \right|^2}{L}.
\end{equation}
We shall use $	\text{PMEPR}(\mathbf{x})$ and $	\text{PMEPR}(\psi(\mathbf{x}))$ invariably whenever the context is clear.
Similarly, for a set $A= \{\mathbf{x}_0, \mathbf{x}_1, \dots, \mathbf{x}_{M-1}\}$ of sequences PMEPR can be defined as
\begin{equation}
	\text{PMEPR}(A)=  \max \{ \text{PMEPR}(\mathbf{x}_i): i=0,1, \dots, M-1 \}.
\end{equation}

\subsection{Bound for PMEPR}

In this subsection, we calculate the PMEPR of the constructed MSCS. First we state a result regarding the PMEPR which is available in the literature.

\begin{lemma}[\cite{xin}]
	If $\mathbf{x}$ is a sequence of a $(2,L,S)$-MSCS, then $\text{PMEPR} (\mathbf{x})$ is upper bounded by $2 S$.
\end{lemma}
A similar statement can be made for a $(M,L,S)$-MSCS using methods similar to \cite{chenMSCS}, which we provide in the following lemma.
\begin{lemma}\label{lemma_PMEPR}
		If $\mathbf{x}$ is a sequence of a $(M,L,S)$-MSCS, then $\text{PMEPR} (\mathbf{x})$ is upper bounded by $M S$.
\end{lemma}

\begin{IEEEproof}
	We briefly sketch the proof here. Let $\{\mathbf{a}_0, \mathbf{a}_1, \dots, \mathbf{a}_{M-1}\}$ be an $(M,L,S)$-MSCS, where $\mathbf{a}_i= \left( a_{i,0} a_{i,1}, \dots, a_{i,L-1}\right)$ for all $i$. We let $\zeta= \exp\left( \frac{2\pi \sqrt{-1}} {S} \right)$ be the $S$-th primitive root of $1$. We define the set of sequences $\mathbf{a}_i^u= \left( a_{i,0} \zeta^{0u}, a_{i,1} \zeta^{1u}, \dots, a_{i,(L-1)} \zeta^{(L-1)u}\right)$ for $u=0,1,\dots, S-1$ and $i=0,1,\dots, M-1$. Then we have
	\begin{equation}
		\begin{split}
			\sum_{u=0}^{S-1} \left|P_{\mathbf{a}_i^u} (t)\right|^2= &\sum_{u=0}^{S-1} \left|  \sum_{k=0}^{L-1} a_{i,k}  \zeta^{ku} z^{f_k} \right|^2\\
			=& LS + 2 \mathcal{R} \left( \sum_{k=0}^{L-1} \rho(\mathbf{a}_i) (k) z^{f_k} \sum_{u=0}^{S-1}  \zeta^{ku} \right),
		\end{split}
	\end{equation}
where $z= \exp\left( 2 \pi t \sqrt{-1}\right)$ and $\mathcal{R}(\cdot)$ is the real part of a complex number. As $\sum_{u=0}^{S-1} \zeta^{ku}$ equals $0$, when $k \!\!\!\mod S \neq 0$ and it equals $S$, otherwise, we have
\begin{equation}
	\begin{split}
		\sum_{i=0}^{M-1} \sum_{u=0}^{S-1} \left|P_{\mathbf{a}_i^u} (t)\right|^2=& MLS.
	\end{split}
\end{equation}
Hence, $\text{PMEPR}(\mathbf{a}_i^u) \leq MS$. For $u=0$, we have $\mathbf{a}_i^u= \mathbf{a}_i$ and it implies $\text{PMEPR}(\mathbf{a}_i)$ is upper bounded by $MS$.
\end{IEEEproof}

\begin{figure}[ht]
	\includegraphics[scale=0.57]{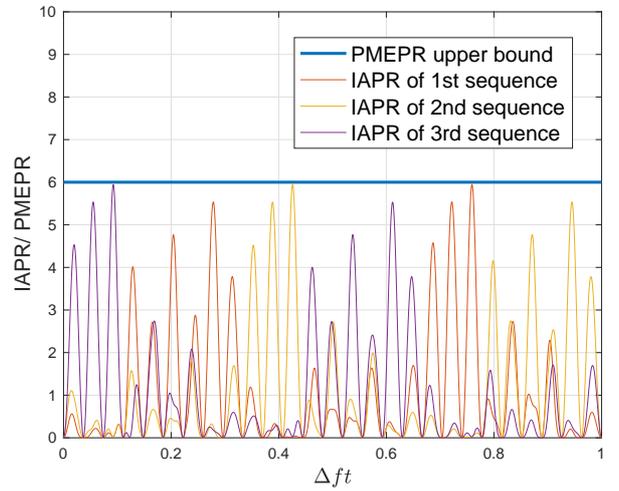}
	\caption{IAPRs of constituent sequences of MSCS having length $54$}
	\label{fig_pmepr}
\end{figure}

\begin{example}\label{eg_pmepr}
	Let $k=2$, $p_1=3$, $p_2=2$, $m_1=3$, $m_2=1$, $s_1=1$. Then, using \textit{Theorem \ref{thm_gen_2}}, we can directly construct a $(3, 54, 2)$-MSCS which has not been reported before. We take $f: \mathbb{Z}_3^3 \times \mathbb{Z}_2 \rightarrow \mathbb{Z}_6$, where  $f_1= 2(v_{3,2} v_{3,3} + v_{3,3} v_{3,1})+ 2 v_{3,1} + 5 v_{3,2} +  v_{3,3}$ and $f_2= 3 v_{2,1}$. In Fig. \ref{fig_pmepr}, we have shown plot of IAPR with respect to $\Delta f t$ for all sequences of the MSCS. The PMEPR value is 5.9465 which agree to the theoretical upper bound $6$ from \textit{Lemma
	\ref{lemma_PMEPR}}.
\end{example}

%

\subsection{Comparison with existing works}
In the TABLE \ref{table_com}, we have compared the proposed construction with the existing ones \cite{chenMSCS, MSCS2, ChaoMSCS} with respect to their parameters. It can be seen from TABLE \ref{table_com} that the proposed constructions provide flexible lengths and set sizes compared to the existing constructions.

\section{Conclusion}
In this paper, we have proposed new constructions of MSCSs, which can be used as alternatives to the conventional GCSs in OFDM. Although there are several direct constructions of MSCSs in the literature, the sequence lengths and set sizes are limited in those. The proposed constructions provide flexible as well as arbitrary sequence lengths and flexible set sizes. For $k=1$, one of the constructed MSCS reduces to type-II ZCS, which has applications in wideband wireless communication systems.




\balance

\bibliographystyle{IEEEtran}

\bibliography{Abhishek_MSCS_ref}


\end{document}